# Recommendation System in Advertising and Streaming Media: Unsupervised Data Enhancement Sequence Suggestions

Kowei Shih[1,4], Yi Han[2,5], Li Tan[3,6]

[1]Independent Researcher, Shenzhen, 518000, China
[2] Meta Fintech, Menlo Park, CA, USA
[3] Fuqua School of Business, Duke University, Durham, NC, USA

[4] skw19@tsinghua.org.cn
[5] han.y@wustl.edu
[6] litan@alumni.duke.edu

**Abstract.** Sequential recommendation is an extensively explored approach to capturing users' evolving preferences based on past interactions, aimed at predicting their next likely choice. Despite significant advancements in this domain, including methods based on RNNs and self-attention, challenges like limited supervised signals and noisy data caused by unintentional clicks persist. To address these challenges, some studies have incorporated unsupervised learning by leveraging local item contexts within individual sequences. However, these methods often overlook the intricate associations between items across multiple sequences and are susceptible to noise in item co-occurrence patterns. In this context, we introduce a novel framework, Global Unsupervised Data-Augmentation (UDA4SR), which adopts a graph contrastive learning perspective to generate more robust item embeddings for sequential recommendation. Our approach begins by integrating Generative Adversarial Networks (GANs) for data augmentation, which serves as the first step to enhance the diversity and richness of the training data. Then, we build a Global Item Relationship Graph (GIG) based on all user interaction sequences. Subsequently, we employ graph contrastive learning on the refined graph to enhance item embeddings by capturing complex global associations. To model users' dynamic and diverse interests more effectively, we enhance the CapsNet module with a novel target-attention mechanism. Extensive experiments show that UDA4SR significantly outperforms state-of-the-art approaches.

# 1. Introduction

With the rapid development of the Internet, recommendation systems have be- come essential on online platforms, effectively capturing users' preferences and addressing the problem of information overload. Among these, sequential recom- mendation (SR), which predicts the next item for users by treating historical in- teractions as temporally ordered sequences, has become a key task widely applied across various scenarios, drawing considerable attention from both academia and industry.

In recent years, significant advancements in sequential recommendation (SR) have been made. Early methods relied on Markov chains [1] to model pairwise transitions between items. To leverage the non-linearity of deep learning, later approaches incorporated various neural architectures. For example, Caser [2] treats user behavior sequences as images, using convolutional neural networks to learn user representations. GRU4Rec and GRU4Rec+ [3, 4] apply Gated Recurrent Units to model entire sessions for better recommendations. To account for the varying importance of items, attention mechanisms were introduced, as seen in SASRec [5] and BERT4Rec [6], to assign weights to items in a sequence. Additionally, graph neural networks have been used in SR to capture complex item relationships. Notable examples include SR-GNN [7] and GC-SAN [8], which use Gated GNNs to model intricate item transitions in session-based contexts.

However, these methods face challenges like data sparsity and noise, which limit their ability to manage scarce training signals and complex item associations. Contrastive learning has emerged as a powerful technique for representation learning and has been applied to SR. For instance, S3Rec [9] uses four auxiliary self-supervised objectives to enhance data representation through mutual information maximization. CL4SRec [10], the first contrastive learning-based SR method, applies augmentations such as item cropping, masking, and reordering to boost performance.

Despite the success of these methods, there are still areas for improvement. Many sequential models focus on the local context within individual sequences, where item co-occurrence is often affected by noise, and the relationships across sequences are not fully captured. Additionally, item popularity follows a long-tail distribution, leading to sparse interactions for many items, which introduces selection bias in their representation. Lastly, although recent works have applied contrastive learning to address data sparsity, they often rely on random augmentations without considering how to tailor contrastive strategies to the unique characteristics of sequential recommendation tasks.

To address these issues, we propose a novel approach by introducing graph contrastive learning into the sequential recommendation to capture informative item representations, laying a solid foundation for accurately modeling users' interests. Specifically, we introduce a framework called UDA4SR, which stands for Unsupervised Data-Augmentation for Sequential Recommendation.

We propose a novel framework for sequential recommendation using graph contrastive learning. First, we construct a Global Item Relationship Graph (GIG) from all interaction sequences. Defining the edges in this graph to properly capture item relationships is challenging, so we quantify adjacency information of various orders as edge weights and apply a threshold to reduce noise. To mitigate popularity bias, we remove more edges for highly popular items. We then apply graph contrastive learning to exploit global associations and enhance item representations. However, designing effective augmentation schemes is difficult since arbitrary methods may distort the graph's underlying semantic information. Inspired by prior work, we use probability-based neighborhood sampling to extract meaningful sub-graphs for augmentation. We address the challenge of embedding augmentation in sequential recommendation through graph contrastive learning, aiming to create a global view and design effective contrastive strategies to mitigate data sparsity and noise issues.

We introduce UDA4SR, a global contrastive data-augmentation framework, where graph contrastive learning is applied on a constructed global graph to capture complex associations between items across sequences.

# 2. Related Work

*2.1 Sequential Recommendation*

Sequential recommendation (SR) aims to predict the next item based on historical interaction sequences. Traditional models, such as those based on Markov chains, focus on capturing short-term item transitions but often overlook long- term user preferences [1]. With the rise of deep learning, several models based on neural networks have been introduced to address this issue.

Recurrent Neural Networks (RNNs) treat users' behavior sequences as time- series data and employ multi-layer GRU structures to capture sequential pat- terns. Notable examples include GRU4Rec and GRU4Rec+ [3, 4]. Additionally, Convolutional Neural Networks (CNNs) and their variants have been used to enhance the performance of sequential recommendation [2, 11]. For instance, Caser [2] utilizes both horizontal and vertical convolutional filters to extract sequential behavior patterns from users.

The attention mechanism has also emerged as a powerful tool in sequential recommendation. SASRec [12] is one of the first frameworks to apply self-attention networks to learn the importance of items within a sequence. BERT4Rec [6]further extends this by modeling bidirectional correlations between sequential items.

In recent years, the use of Graph Neural Networks (GNNs) has gained trac- tion due to their ability to capture complex item transition patterns within user sequences. Several studies have applied GNNs to SR. Notably, SR-GNN [7] was one of the first to use gated GNNs for information propagation, captur- ing complex item transitions in session scenarios. GC-SAN [8] combines GNNs with self-attention mechanisms to model both local and long-range dependencies between items.

However, many of the aforementioned methods rely primarily on prediction loss to optimize the representation of the entire sequence, improving recommen- dation performance but often neglecting valuable unsupervised signals.

### 3. Preliminary
*3.1Global Item Relationship Graph*
Unlike previous methods [2, 5] that focus on the local context within individual sequences, we construct a Global Item Relationship Graph (GIG) from all interaction sequences, connecting items with undirected edges. Repeated items within a sequence are not removed. Edge weights reflect the adjacency of items across sequences. Specifically, an n-GIG is defined where the weight of an edge between items is the sum of their direct connections at various intervals across all sequences. This setting, inspired by prior work, is used to compute edge weights, which are then normalized.

### 4. Methodology
In this section, we introduce the proposed model, UDA4SR, which consists of three main components: Data-Augmentation, Multi-Interest Extraction and Fusion, and Multi-Task Learning.

*4.1 Data-Augmentation*
To address data sparsity and enhance diversity in recommendation systems, we propose an independent GAN-based data augmentation framework. The generator produces realistic user-item interaction sequences from partial input, while the discriminator assesses authenticity. Adversarial training optimizes both components, incorporating a diversity-promoting term to prevent mode collapse. The trained generator creates synthetic sequences, enriching real data for robust recommendation model training. Additionally, Graph Contrastive Learning (GCL) enhances representation learning in graph-based systems by integrating GAN-generated data while maintaining semantic consistency through careful augmentation strategies.

*4.2Multi-Interest Extraction and Fusion*
In advertising and streaming media recommendation systems, user interests are often diverse and dynamic. Accurately modeling these preferences is crucial for improving recommendation effectiveness. Previous studies have shown that Capsule Networks (CapsNet) excel in capturing user interests and underlying patterns. Building upon this, we propose a multi-interest extraction and fusion approach that integrates CapsNet with a target-attention mechanism to enhance user preference modeling, thereby improving recommendation accuracy.

User interactions are represented as a temporal sequence of item embeddings. To effectively capture sequential dependencies and long-term interest patterns, we employ a Transformer model to encode user behavior. This structure allows for global context modeling while incorporating residual connections to refine representation learning.

To extract multiple interests, we define a set of interest capsules, each aggregating information from user interactions. These capsules iteratively update their representations using dynamic routing, refining interest vectors through nonlinear transformations. A target-attention mechanism is then applied to align these interest representations with a given item, generating a personalized user preference vector. The final recommendation score is computed based on the interaction between this preference vector and the target item embedding, ensuring a more accurate and adaptive recommendation strategy for advertising and streaming media.

*4.3 Multi-Task Learning*

Additionally, the framework incorporates multi-task learning by jointly optimizing a prediction task and a contrastive learning task. The contrastive learning task distinguishes augmented representations of the same item from others, enhancing the alignment of positive examples and separation of negatives, using a contrastive loss function. The prediction task leverages binary cross-entropy loss to refine the recommendation accuracy. A final objective function combines these losses with regularization to balance predictive performance and prevent overfitting. Parameters are optimized using gradient descent.

## 5. Conclusion

*5.1 Experimental Settings*

We evaluated the model on four public datasets across different domains to assess scalability and adaptability. To maintain data quality, users and items with fewer than 15 interactions were filtered, and interaction sequences were sorted by timestamp, with a maximum length of 50. In advertising and social media recommendations, preprocessing is crucial due to varying user behavior patterns, such as high-frequency interactions on social media and the need to capture long-tail products in advertising.

User interaction sequences were split into training (80%), validation (10%), and testing (10%), ensuring effective data utilization while preserving authenticity, particularly for rapidly shifting preferences. To ensure a balanced evaluation, datasets included both frequent and sparse interactions.

We used Recall@k and NDCG@k (k=10,20) for performance comparison. A full-ranking strategy was applied to reduce sampling bias, making it especially effective in assessing long-tail interest modeling for ad placement and content ranking.

*5.2 Performance Comparison*

In this section, we evaluate the recommendation performance using Recall and NDCG metrics across four public datasets, as shown in Table 2, and derive the following observations:

**Table 2.** The performance of different models.

| Dataset | Metric | GRU4Rec | Caser | SASRec | SRGNN | GCSAN | S3Rec | UDA4SR |
|---|---|---|---|---|---|---|---|---|
| ML-1M | R@10 | 0.184 | 0.167 | 0.193 | 0.188 | 0.194 | 0.199 | 0.215 |
|  | N@10 | 0.246 | 0.231 | 0.260 | 0.250 | 0.257 | 0.263 | 0.297 |
|  | R@20 | 0.273 | 0.254 | 0.289 | 0.276 | 0.285 | 0.294 | 0.321 |
|  | N@20 | 0.256 | 0.243 | 0.273 | 0.262 | 0.270 | 0.279 | 0.295 |
| Sports | R@10 | 0.104 | 0.101 | 0.114 | 0.113 | 0.117 | 0.122 | 0.134 |
|  | N@10 | 0.089 | 0.081 | 0.099 | 0.093 | 0.098 | 0.104 | 0.113 |
|  | R@20 | 0.130 | 0.113 | 0.140 | 0.133 | 0.138 | 0.143 | 0.162 |
|  | N@20 | 0.099 | 0.093 | 0.109 | 0.121 | 0.106 | 0.112 | 0.123 |
| Yelp | R@10 | 0.064 | 0.064 | 0.081 | 0.083 | 0.080 | 0.088 | 0.099 |
|  | N@10 | 0.048 | 0.044 | 0.060 | 0.053 | 0.061 | 0.068 | 0.075 |
|  | R@20 | 0.104 | 0.104 | 0.120 | 0.123 | 0.129 | 0.134 | 0.145 |
|       | N@20 | 0.059 | 0.058 | 0.071 | 0.064 | 0.072 | 0.079 | 0.086 |
| Books | R@10 | 0.068 | 0.063 | 0.082 | 0.076 | 0.084 | 0.093 | 0.104 |
|       | N@10 | 0.046 | 0.045 | 0.058 | 0.050 | 0.058 | 0.067 | 0.076 |
|       | R@20 | 0.103 | 0.101 | 0.124 | 0.111 | 0.120 | 0.129 | 0.139 |
|       | N@20 | 0.054 | 0.055 | 0.070 | 0.060 | 0.069 | 0.078 | 0.087 |

First, comparing graph neural network (GNN)-based methods with traditional ones such as BPRMF, GRU4Rec, and Caser, SR-GNN and GC-SAN exhibit significantly better performance. This validates existing studies that show GNN methods are superior in capturing high-order interactions between users and items, thereby improving recommendation accuracy.

Second, SASRec, which integrates a self-attention mechanism, outperforms GRU4Rec and Caser. This highlights the notable advantage of self-attention mechanisms in sequence modeling, particularly in capturing long-term dependencies between items, which is crucial in scenarios where user behaviors exhibit strong temporal characteristics.

Finally, among self-supervised learning methods, CL4Rec and S3Rec surpass baseline methods that use single-paradigm losses, demonstrating the benefits of incorporating self-supervised tasks into sequential recommendation models. However, their augmentation strategies primarily focus on local contexts within individual sequences, limiting their ability to leverage intricate item associations across sequences, thus constraining further performance improvements.

To address these limitations, particularly in the context of advertising and social media recommendation systems, this study proposes a novel framework, UDA4SR, tailored for unsupervised data enhancement in sequence recommendation. The framework leverages Generative Adversarial Networks (GANs) for data augmentation, enriching the diversity and richness of training data. Subsequently, a Global Item Relationship Graph (GIG) is constructed to capture complex global associations across user interaction sequences. Graph contrastive learning is applied to enhance item embeddings. Moreover, the CapsNet module is refined with a target-attention mechanism to model users' dynamic and diverse preferences more effectively.

UDA4SR demonstrates strong adaptability in complex environments, such as personalized ad delivery and social media content recommendations, significantly addressing challenges like limited supervised signals and data noise. Experimental results reveal that UDA4SR substantially improves recommendation accuracy and robustness, providing a promising direction for optimizing recommendation systems in advertising and streaming media contexts. Future research could focus on enhancing cross-sequence modeling and integrating multi-source data to further advance intelligent applications in these fields.

## 6. Conclusion

This study introduces UDA4SR, an unsupervised data enhancement sequential recommendation method tailored for advertising and streaming media recommendation systems. UDA4SR is an innovative graph contrastive learning approach designed to capture potential relevance within both local and global item contexts. First, we construct a Global Item Relationship Graph (GIG) based on all user interaction sequences and apply graph neural networks (GNNs) to sub-graphs sampled from neighborhoods. This enhances item representations with complex associative information. Simultaneously, to better model the diverse preferences of users in advertising and social media recommendations, we extend CapsNet with a target-attention mechanism and derive final recommendations through a multi-task framework.

The effectiveness and superiority of this method are validated through extensive experiments on four public datasets. Compared to existing approaches, UDA4SR achieves significant improvements in recommendation accuracy, particularly in dynamic and highly personalized recommendation tasks in advertising and streaming media. This method enables advertisers to target audiences more precisely and enhances social media platforms' ability to capture user interests, optimizing both advertising efficiency and user experience.

Future work will focus on integrating comparison information at different granularities to address challenges related to data sparsity and noise in sequential recommendations. Additionally, we plan to explore extending the model to broader recommendation scenarios, such as real-time ad placement strategy optimization and social media content recommendations, providing smarter decision support for relevant industries.